\documentclass[%
aip,reprint,amsmath,amssymb,cha,
]{revtex4-1}

\ifx\pdfoutput\undefined
\usepackage[dvips]{graphicx}
\graphicspath{{figures/}}
\else
\usepackage[pdftex]{graphicx}
\graphicspath{{figures/}}
\usepackage[pdftex]{hyperref}
\fi
\newcommand{\tobs}{T_{\textrm{obs}}}

\begin{document}

   \title{Effect of discrete time observations on synchronization in Chua
   model and applications to data assimilation}

\author{Md. Nurujjaman}
\email{jaman@math.tifrbng.res.in}
\author{Sumanth Shivamurthy}
\author{Amit Apte}
\email{apte@math.tifrbng.res.in}
\affiliation{Centre for Applicable Mathematics, Tata Institute of Fundamental
Research, Sharada Nagar,
Chikkabommasandra, Bangalore 560065, India}
\author{Tanu Singla}
\author{P. Parmananda}
\affiliation{Department of Physics, Indian Institute of Technology Bombay, Powai, Mumbai 400076, India}

\date{\today}

\begin{abstract}
Recent studies show indication of the effectiveness of synchronization
as a data assimilation tool for small or meso-scale forecast when
less number of variables are observed frequently. Our main aim here is
to understand the effects of changing observational frequency and
observational noise on synchronization and prediction in a low
dimensional chaotic system, namely the Chua circuit model. We perform
{\it identical twin experiments} in order to study synchronization
using discrete-in-time observations generated from independent model
run and coupled unidirectionally to the model through $x$, $y$ and $z$
separately. We observe synchrony in a finite range of coupling
constant when coupling the $x$ and $y$ variables of the Chua model but
not when coupling the $z$ variable. This range of coupling constant
decreases with increasing levels of noise in the observations. The
Chua system does not show synchrony when the time gap between
observations is greater than about one-seventh of the Lyapunov time.
Finally, we also note that prediction errors are much
larger when noisy observations are used than when using observations
without noise.
\end{abstract}

\maketitle
\begin{quotation}

The possibility of using chaotic synchronization for
data assimilation is entertained. In particular,
synchronization observed for two chaotic Chua models
under unidirectional coupling is shown to be of relevance for
forecasting using data assimilation techniques.
Our results indicate that prediction method
employed in the present is extremely sensitive
to observational frequency, Lyapunov exponent and
the phase at which the coupling is switched off.
\end{quotation}

\section{\label{sec:intr} Introduction}
Data assimilation (DA) is a powerful and versatile method for
combining partial, noisy observational data of a system with its
dynamical model, generally numerically implemented, to generate state
estimates of nonlinear, chaotic systems. A variety of data
assimilation methods, broadly separated into deterministic or
probabilistic methods, have been developed over the past few decades
and used mainly in the earth sciences.\cite{book:lewisvarahan06,
  book:park08, Kalnay03, book:malanotte96, Ben01, book:evensen,
  PhysicaD:Ide} On the other hand, when two or more chaotic systems
are coupled, they may adjust the properties of their motion due to
coupling in such a way that their evolution becomes
\emph{synchronized}. Several types of coupling, unidirectional,
mutual, common driving, etc., and their effects in the form of a
variety of synchronizations such as identical, generalized, phase,
lag, amplitude, etc., as well as applications of synchronizations have
been and continue to be a very active area of
research.\cite{prl:pecora, book:pikovosky}

We will be interested in this paper with \emph{sequential data
  assimilation}, which incorporates observations sequentially in
time. This method can be thought of as synchronization through
unidirectional coupling, with observations acting as the master system
and the numerical model acting as the slave system.\cite{prl:pecora,
  atmos:yang, npg:Duane, JGR:Szendro} In fact, this is exactly the
nudging method for assimilation, suggested first in
meteorology,\cite{mwr:hoke} and later used for a number of different
studies.\cite{jgr:verron, tellus:stauffer, mwr:bao, jaot:wang,
  ngp:aurox} Recent studies show that such techniques can also be used
for assimilating data in order to track complex spatio-temporal
dynamics of excitable media,\cite{chaos:berg} and forecast the state
of a time-delayed high-dimensional system, e.g., in chaotic
communication.\cite{prl:cohen} Synchronization based DA method may be
useful for small or mesoscale predictions, specially when observations
are frequent and coupled to a small number of
variables.\cite{atmos:yang,ngp:aurox} This method is relatively fast
and robust, and can be implemented easily.\cite{Oceanography:Stammer}

Two of the main characteristics of the observational data used in
earth sciences applications is that they are discrete in time and are
noisy. The main aim of this paper is to study synchronization with
these characteristics in mind, as we explain below.
\begin{enumerate}
\item It has been observed recently\cite{DynAtmOcean:ide,
  jhydro:walker, waterres:walker, tellus:apte} that the time period
  between two observations is one of the important factors which
  significantly affect conventional DA schemes. Though frequent
  updates may produce better results depending on the quality of the
  model as well as the observations, they increase the burden of
  computational cost.\cite{advspaceres:Riishojgaard} Since the choice
  of observational frequency depends on many factors such as cost and
  ease of observations, it is important to understand in detail how
  the time period between observations affects the accuracy of
  predictions which in turn depends on the accuracy of the data
  assimilation.

  For these reasons, we investigate the effect of varying
  observational frequency, i.e., the gaps between the observations, on
  discrete time synchronization -- this discussion forms the first
  part of the paper. As expected, we find that observations which are
  farther apart lead to poorer synchronization and consequently low
  accuracy of prediction. We also find that there is a certain
  threshold, which in the system we study is roughly of the order of
  the inverse of Lyapunov exponent of the system, such that using
  observations with a time period greater than that threshold leads to
  no synchronization.

\item It is of course well known that the observational noise plays
  crucial role in data assimilation as well as in
  synchronization.\cite{PRL:Neiman, PRL:Teramae, PRE:Guan} Thus we
  also carry out the study described in the above paragraph with noisy
  observations with different noise levels, in order to investigate
  the effect of noise on synchronization. The discussion of these
  results forms the second part of the paper. Again as expected, the
  increase in noise leads to poorer synchronization and prediction.
\end{enumerate}
We present these results in detail in section~\ref{sec:results} and a
discussion of these results along with directions for further studies
are presented in section~\ref{sec:conclude}.

In order to investigate the effects of these two aspects, we perform
the so-called ``identical twin experiments''\cite{tellus:talagrand81}
as follows. We generate observations using a known trajectory, called
the ``true'' trajectory of a numerical model. We use these
observations (master) to try to synchronize the same numerical model
starting with a different initial condition (slave). We quantify the
``degree of synchronization'' by calculating the root mean square
(RMS) error of the slave trajectory with respect to the true
trajectory, leaving out the first half of the time-span, in order to
get rid of transients. We assess the effect on predictions by
calculating the RMS error over certain time-spans after the
synchronization is stopped and the slave model is let to run by
itself. Currently we are investigating the effect of these two
factors, the time period between observations and the noise level, on
data assimilation methods such as ensemble Kalman filter
(ENKF).\cite{book:evensen}

The dynamical model we use is that of a chaotic Chua
circuit.\cite{IJBC:chua, pre:singla} One of the main reason for this
choice is that in future we would like to use the actual data from the
circuit experiments being currently performed by two of the
authors.\cite{pre:singla} This model is described in detail in the
following section.

\section{Chua circuit model and synchronization}
\label{sec:model}

To study synchronization we have used the Chua model used in
Ref.~\onlinecite{pre:singla}. The model of the circuit is made
dimensionless by substituting $\tau=\beta t$, where
$\beta=(R_1C_1)^{-1}$, and $R_1$ and $C_1$ are the resistance and
capacitor used in the circuit respectively. The dimensionless model is
given by.
\begin{eqnarray}
\frac{dx}{d\tau}&=&\frac{1200}{R} (y-x)- g(x) \,,\nonumber \\
\frac{dy}{d\tau}&=&\sigma\left[\frac{1200}{R} (y-x) +z\right] \,,\nonumber\\
\frac{dz}{d\tau}& =&-c [y +r'z] \,,\label{eqn:master}
\end{eqnarray}
where
$$ g(x) = \left\{ \begin{array}{ll}
-x &\mbox{  $|x|<1$} \\
-[1+b(|x| -1)]sign(x) &\mbox{$1<|x|\le10$}\\
10[(|x| -10)-(9b+1)]sign(x) &\mbox{$10<|x|$}
      \end{array} \right. $$ 
and $\sigma, c, r',$ and $b$ are the parameters whose values depend on
the capacitors, resistances and inductor used in the
circuit.\cite{pre:singla} The resistance $R$ is the control parameter
of the system. Depending on R, the system shows chaotic or regular
behavior. In the present study, we have chosen $\sigma = 4.6/69
\approx 0.067$, $c = (1200 \times 3300 \times 4.6/8.5) \times 10^{-6}
\approx 0.779$, $r' = 85/1200 \approx 0.071$ and $b = 1 - 1200/3300
\approx 0.636$. The model was integrated using the fourth order
Runge-Kutta numerical scheme with integration step $\Delta \tau =
0.01$, and the output of the model has been expressed in terms of real
time $t = \tau/\beta$ rather than in terms of the dimensionless time
$\tau$.
\begin{figure}
\includegraphics[width=8.5 cm]{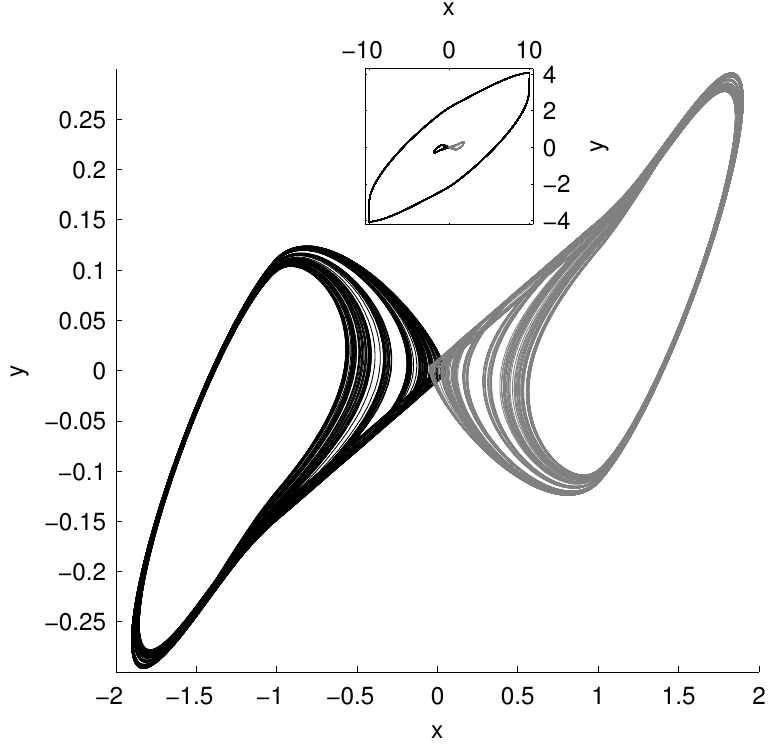}
\caption{Two Chua attractors for the system in
  Eqs.\eqref{eqn:master}. Inset shows the large attractor, which
  represents the diverging dynamics, along with the above two
  attractors.}
\label{fig:attractors}
\end{figure}

The model equations show three attractors depending upon the initial
conditions as shown in Fig.~\ref{fig:attractors}(inset), and zoomed
version of the smaller attractors by black and gray plots. As the
basin of attractors is not known, it is not possible to tell \emph{a
  priory} which initial condition will lead to a particular
attractor. Generally, for high values of initial condition
(approximately greater than $1$ in absolute value), system goes to the
large limit cycle attractor [inset of Fig.~\ref{fig:attractors}], and
for smaller initial conditions, system stays on any of the two small
attractors [Fig.~\ref{fig:attractors}]. As the larger attractor is not
of experimental interest, the dynamics on this attractor is considered
as divergent dynamics.

In the present simulation, we have used $R = 1245 \Omega$ for which
the system shows chaotic behavior. The ``true'' (or master) initial
condition is chosen such that the trajectory stays on any one of the
small attractors depending upon the initial conditions. The largest
Lyapunov exponent ($\lambda$) is positive ($\approx 4.04$ bits/msec),
and corresponding Lyapunov time is $t_{\lambda} = 3.6 \times 10^{-4}
sec$. $\lambda$ was estimated using Rosenstein
techniques~\cite{physicaD:Rosenstein} and is consistent with the
earlier results.\cite{JCktSystComp:parlitz}

In order to generate any one set of observations, we first simulated a
long trajectory on one of the smaller attractors.  We refer to such a
trajectory as ``truth'' in keeping with other data assimilation
literature. Next, we sub-sampled this trajectory to extract several
sets of discrete-time equispaced observations with varying time
interval between the observations, which will henceforth be denoted by
$T_{\textrm{obs}}$.  For the present study, observations were
generated with four different values of $T_{\textrm{obs}}$, namely,
$2.0\times10^{-5}$, $5.0\times10^{-5}$, $7.0\times10^{-5}$, and
$1.0\times10^{-4}$ sec. Noisy observations were created by adding
noise of particular intensity ($D$) to these discrete-time
observations, where $D$ was chosen to be a specified fraction of the
standard deviation of the attractors. The results presented in
Sec.~\ref{sec:results} are for one specific set of observations, but
the qualitative features of these results were found to be identical
when we used observational sets generated from many different initial
conditions.

\subsection{Master-slave coupling}

The model (slave) was unidirectionally coupled with $N$ discrete-time
observations (master) at times $\tau_1,\dots,\tau_N$ as follows.
\begin{widetext}
\begin{eqnarray}
\frac{dx}{d\tau}&=&\frac{1200}{R} (y-x)- g(x) -
k_x\sum_{i=1}^N \left[x(\tau) - x_{obs}(\tau_i)\right]\delta(\tau-\tau_i) \nonumber \\
\frac{dy}{d\tau}&=&\sigma\left[\frac{1200}{R} (y-x) +z\right] -
k_y\sum_{i=1}^N \left[y(\tau) - y_{obs}(\tau_i)\right]\delta(\tau-\tau_i) \nonumber \\
\frac{dz}{d\tau}& =&-c [y +r'z] -
k_z\sum_{i=1}^N \left[z(\tau) - z_{obs}(\tau_i)\right]\delta(\tau-\tau_i) 
\label{eqn:sync}\end{eqnarray}
\end{widetext}
where, $k_x, k_y$, and $k_z$ represent the $x$, $y$, and $z$
coupling. In the case of continuous time observations which is also
discussed below, the slave is coupled to the master directly, by
replacing the summations of delta functions in the above equations by
terms such as $k_x[x(\tau) - x_{\textrm{obs}}(\tau)]$. We have quantified
synchronization using RMS error of the difference of the synchronized
and true trajectory over the last half of the trajectory, and the
quality of prediction is again quantified by the RMS error with
respect to truth over varying time periods, as discussed in detail
below. The results about synchronization and prediction for different
$T_{\textrm{obs}}$ and noise level ($D$), and a comparison of these
results with the case of continuous time synchronization are presented
in the next section.

\section{Results}
\label{sec:results}

The main aim of this section is to discuss the effects of the
observational time period $\tobs$ and the observational noise variance
$D$ on the root mean square of the difference between the slave system
and the master, both during the time period in which the slave is
coupled to the master (synchronization phase) and after the coupling
is switched off (the prediction phase).

\subsection{Synchronization with observations without noise ($D=0$)}
\label{subsec:obs_time}
\begin{figure}[t]
\includegraphics[width=8.5 cm]{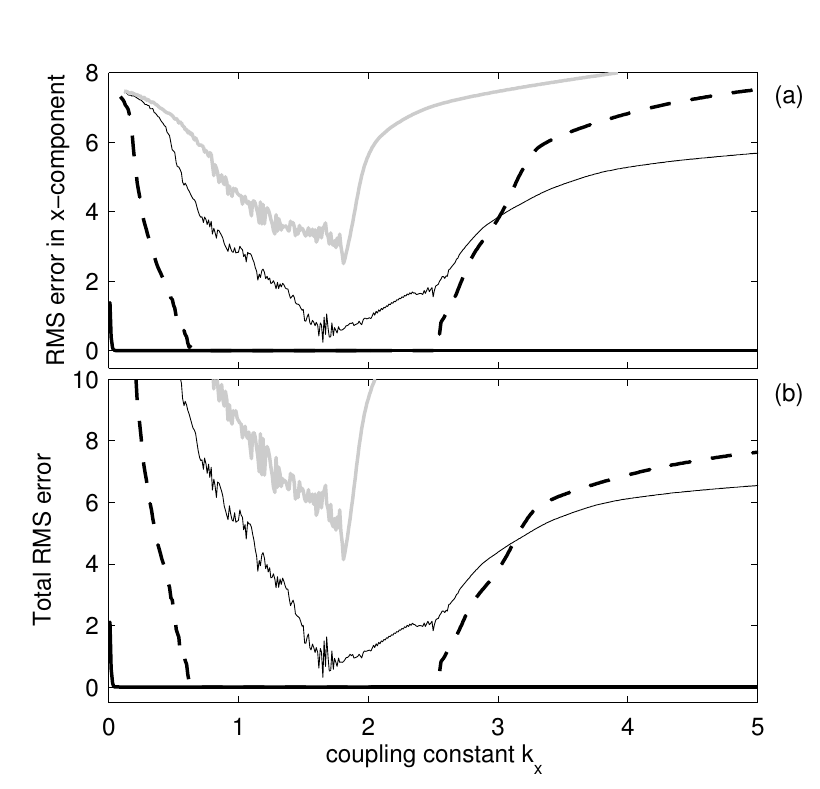}
\caption{RMS errors in the $x$-component (top) and the total RMS error
  (bottom) when $x$-component was observed. In both these plots as
  well as in Fig.~\ref{fig:rms_y}, thick solid line is for continuous
  time synchronization while dashed, thin solid, and gray lines are
  for cases with $T_{\textrm{obs}}=2.0\times10^{-5}$,
  $5.0\times10^{-5}$, and $7.0\times10^{-5}$ sec respectively. The
  $y$- and $z$-components also show similar qualitative behavior. Note
  that RMS error above approximately $0.5$ indicates loss of
  synchronization.}
\label{fig:rms_x}
\end{figure}                

In order to investigate the performance of the synchronization, we
have first studied the case of coupling only the $x$-component,
i.e. setting $k_y = 0 = k_z$ in Eqns.~\ref{eqn:sync} while $k_x$ is
varied from zero to higher values. We quantify synchronization of each
components by calculating the RMS error of the component-wise difference of
master and slave trajectories, and also the total RMS error. For
continuous case, the slave model synchronizes with the master within a
very short time. Solid lines in Fig~\ref{fig:rms_x} (a) and (b) show
the RMS errors in $x$-component and total errors respectively. They
show that the system started synchronizing around $k_x\approx 0.005$,
and achieved full synchronization around $k_x\approx 0.03$ as the RMS
errors are almost zero. The system remained in synchronized state even
at higher values of $k_x$.

\begin{figure}
\includegraphics[width=8.5 cm] {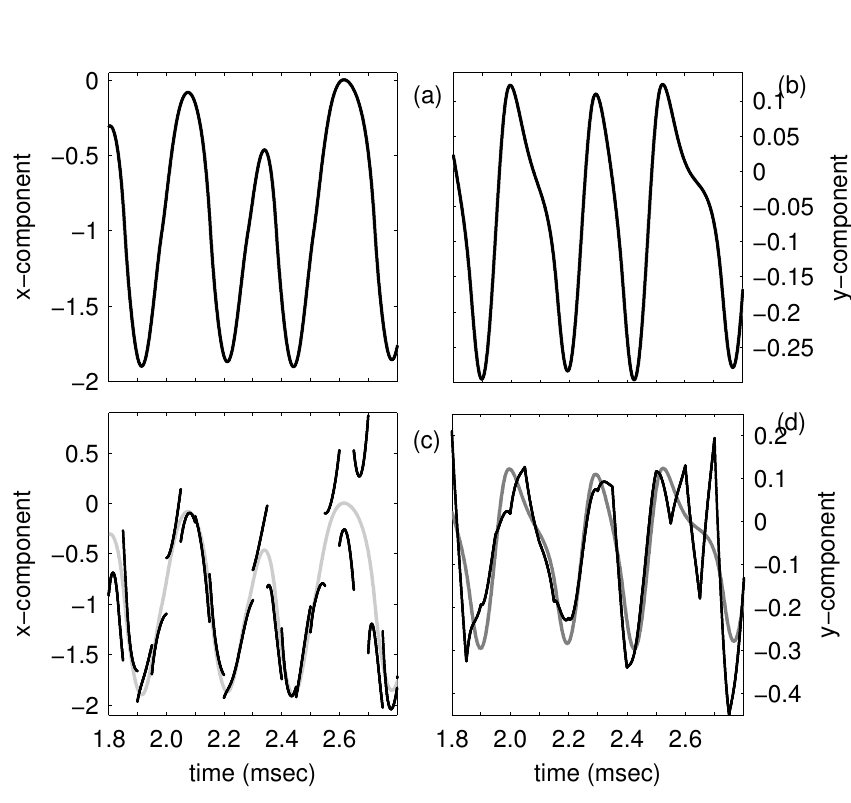}
\caption{First and second columns represent the $x$ and $y$ components
  of the slave trajectories (solid line) and model trajectories (gray
  line) at $k_x = 1.74$ for the observations taken at
  $T_{\textrm{obs}}=2.0\times10^{-5}$ (top) and $5.0\times10^{-5}$
  (bottom). Note that the jumps in $x$-component and in $y$-derivative
  are because of the $\delta$-functions in Eqs.~\eqref{eqn:sync}. }
\label{fig:time_sr}
\end{figure}

Synchronization with $x$-component observations at discrete time is
shown in Fig.~\ref{fig:rms_x}. In this figure, the dashed-lines show
the RMS errors when $x$-observation were recorded with
$T_{\textrm{obs}}=2.0\times10^{-5}$ sec, which is approximately 18
times less than $t_{\lambda}$. The figure also shows that the model
started synchronizing around $k_x=0.18$ with the observations and
achieved full synchronization around $k_x>0.5$.  For $k_x>2.5$ model
diverges from the observations. So the range of the coupling $k_x$ for
which there is synchrony has become finite when observations were
taken at a finite interval. Synchrony is also clear from the plots of
$x$ and $y$ time series of the observational trajectories and slave
trajectories shown in Fig.~\ref{fig:time_sr}.  They show the model
(gray lines) gets perfectly synchronized with the observational
trajectory (black lines).

Note that RMS error above approx. $0.5$ indicates a loss of
synchronization, hence effectively synchronization is obtained only
for the first two of these cases, i.e., continuous time observations
and with $\tobs = 2.0 \times 10^{-5}$ sec.

Black thin solid lines [Fig~\ref{fig:rms_x}] show the RMS errors, when the
observations were taken with $T_{\textrm{obs}}=5.0\times10^{-5}$ sec.~
($ \approx \frac{1}{7} t_{\lambda}$). In this case errors decreased slowly with
$k_x$ and showed partial synchronization only in a very small range of
$k_x$ around $k_x \approx 1.7$. The time series at $k_x=1.74$ in plots
(c) and (d) of Fig.~\ref{fig:time_sr} show such partial synchrony
between the observed trajectory (gray lines) and the slave model
(black line) for $\tobs = 5.0\times10^{-5}$ sec.~The behaviour of the
$z$-component was similar to that of the $y$. For
$T_{\textrm{obs}}=7.0\times10^{-5}$ sec, the RMS error is least around
$k_x=1.19$ but even in this case, it is far from being
synchronized. These results show that $T_{\textrm{obs}}$ is an
important factor in synchronization that must be considered in DA
experiments.
\begin{figure}
\includegraphics[width=8.5 cm]{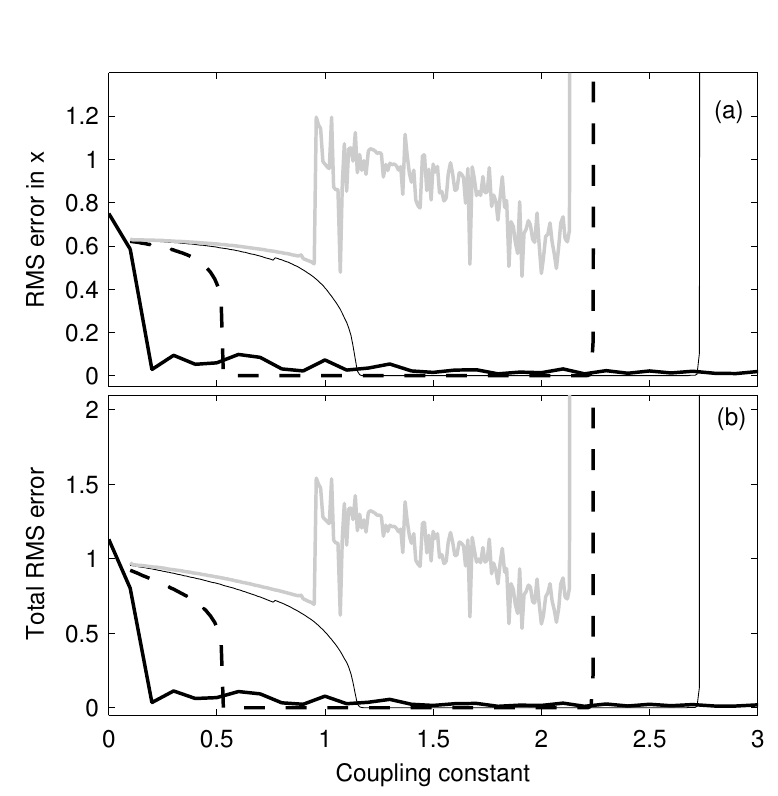}
\caption{RMS errors in the $x$-component (top) and the total RMS error
  (bottom) when $x$-component was observed. The lines are for same
  $\tobs$ as described in Fig.~\ref{fig:rms_x}. Comparing the RMS
  error for $\tobs = 5.0 \times 10^{-5}$ sec (thin solid lines) in
  this figure and in Fig.~\ref{fig:rms_x}, we see that
  $y$-observations are more informative than the $x$-observations.}
\label{fig:rms_y}
\end{figure}

When $y$-component was observed, i.e., when we set $k_x = 0 = k_z$ and
vary $k_y$, the model showed synchrony for different range of
$k_y$ than that was for $x$ observations. In Fig.~\ref{fig:rms_y}, the
solid lines show the RMS errors in (a) $x$-component and (b) total RMS
error for continuous observation case. Around $k_y=0.2$ the model
achieved synchrony and remains in this state even at higher values of
$k_y$. For $T_{\textrm{obs}}=2.0\times10^{-5}$ sec (dashed lines),
slave system achieved full synchronization around $k_y=0.53$, and
remains synchronized for $0.53<k_y<2.22$. When
$T_{\textrm{obs}}=5.0\times10^{-5} sec$ (thin solid), the
synchronization region shifts to $1.14<k_y<2.72$. Similar to
$x$-observation case, there is no synchronization for the
$y$-observation with $T_{\textrm{obs}}=7.0\times10^{-5} sec$ (gray
line) and above. Thus we see that the $y$-component is a more dominant
dynamical variable than the $x$-component.

\begin{figure}
\includegraphics[width=8.5 cm]{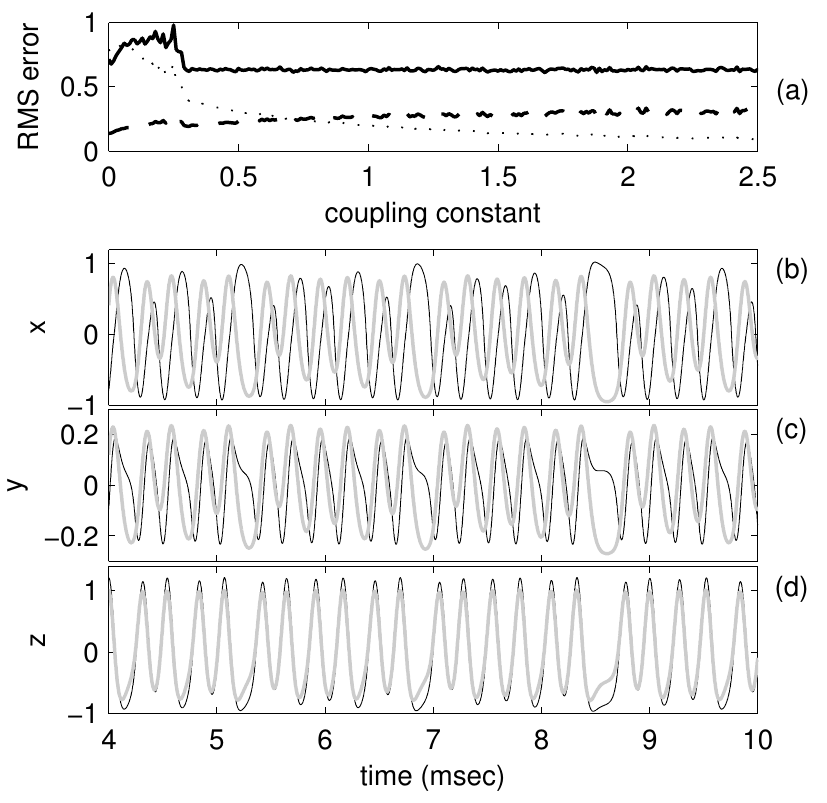}
\caption{(a) RMS errors in $x$ (solid), $y$ (dashed) and $z$-component
  (dotted line) when $z$-component was observed continuously in time,
  showing that there is no synchronization when $z$ is observed. Time
  series of slave (gray) and master (black line) at $k_z= 1.2$ is
  shown in (b) $x$, (c) $y$ and and (d) $z$. The $x$- and
  $y$-components of the slave have been multiplied by factors of 20
  and 0.5 respectively after subtracting mean, in order to see the comparison clearly.}
\label{fig:rms_z}
\end{figure}

\begin{figure}
\includegraphics[width=8.5 cm]{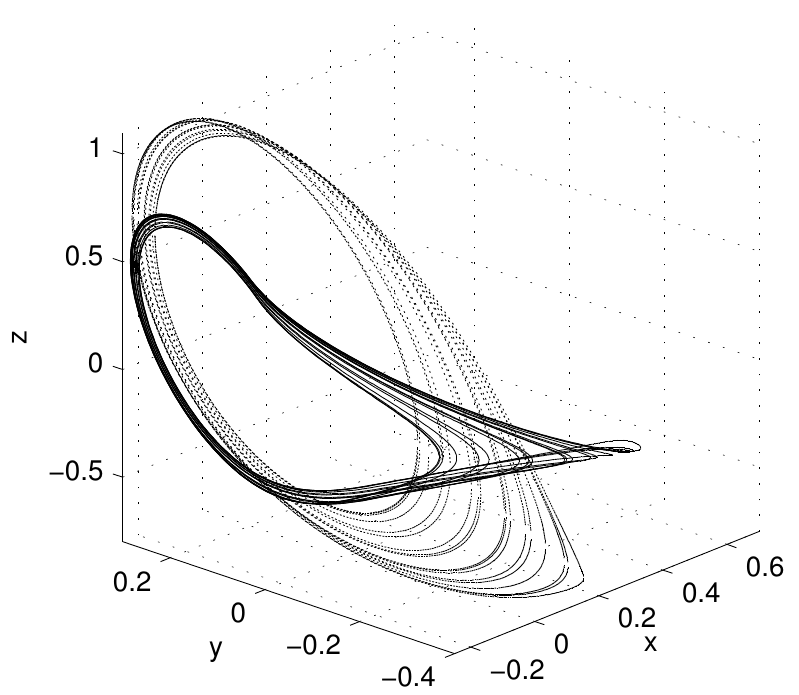}
\caption{Attractors of the master (solid black) and slave system
  (dotted line) when $z$ was observed. The structures of these
  attractors are clearly different from each other.}
\label{fig:z_attract}
\end{figure}
When model was coupled with master through $z$-observations ($k_x = 0
= k_y$), there was no synchronization even for the continuous
observations case. Panel (a) of Fig~\ref{fig:rms_z} shows the RMS
errors in $x$ (solid line), $y$ (dashed line), and $z$ (dotted line)
components as $k_z$ is varied. This is also clear from the plots of
$x$, $y$, and $z$ time series of the master (gray) and slave (solid)
shown in panels (b), (c), and (d), respectively. The difference
between the master (solid line) and slave (dotted line) is also clear
from the attractors shown in Fig~\ref{fig:z_attract}. A possible
explanation for not having synchronization with $z$ coupling is that
this variable may not contain sufficient information about the
dynamics. Hence in the current regime of the Chua system $x$ and $y$
contain useful information for synchronization but not the
$z$-component.\cite{atmos:yang} Finally as the model did not show any
synchrony for continuous case, obviously we did not get any synchrony
with discrete time observations either.

\subsection{Synchronization with noisy observations}
\label{subsec:noisy_obs}

\begin{figure}
\includegraphics[width=8.5 cm]{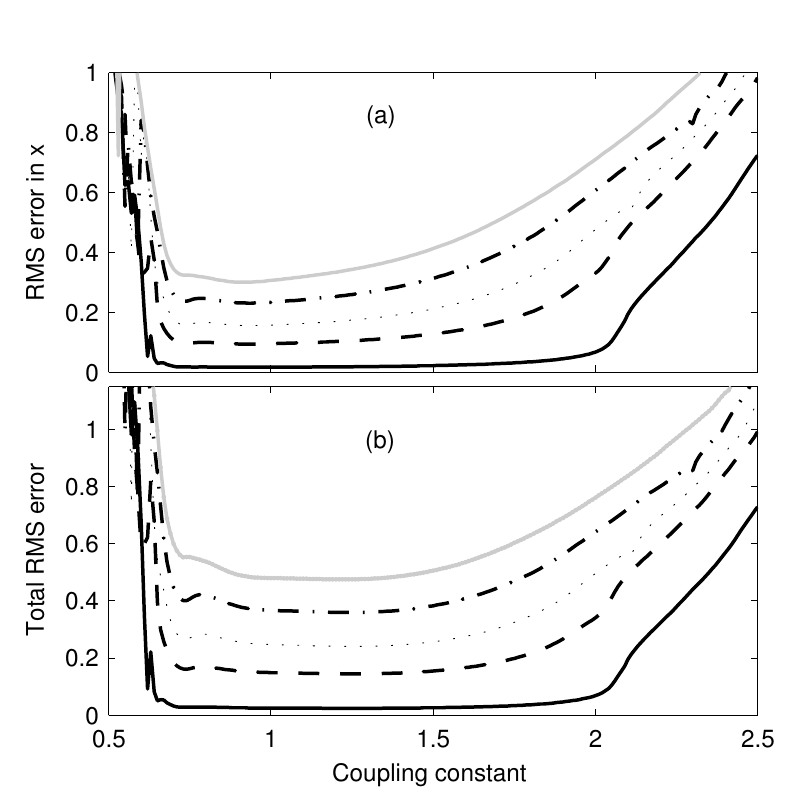}
\caption{RMS error of synchronization when noisy $x$-component
  observations with $T_{\textrm{obs}}=2.0\times10^{-5}$ are coupled to
  the slave model: (a) RMS error in $x$-component and (b) total RMS
  error. The solid, dashed, dotted, dashed-dotted, and gray lines
  represent the RMS errors for noise levels $D=2\%, 10\%, 20\%, 30\%$,
  and $40\%$, respectively.}
\label{fig:rms_nz}
\end{figure}

\begin{figure}
\includegraphics[width=8.5 cm]{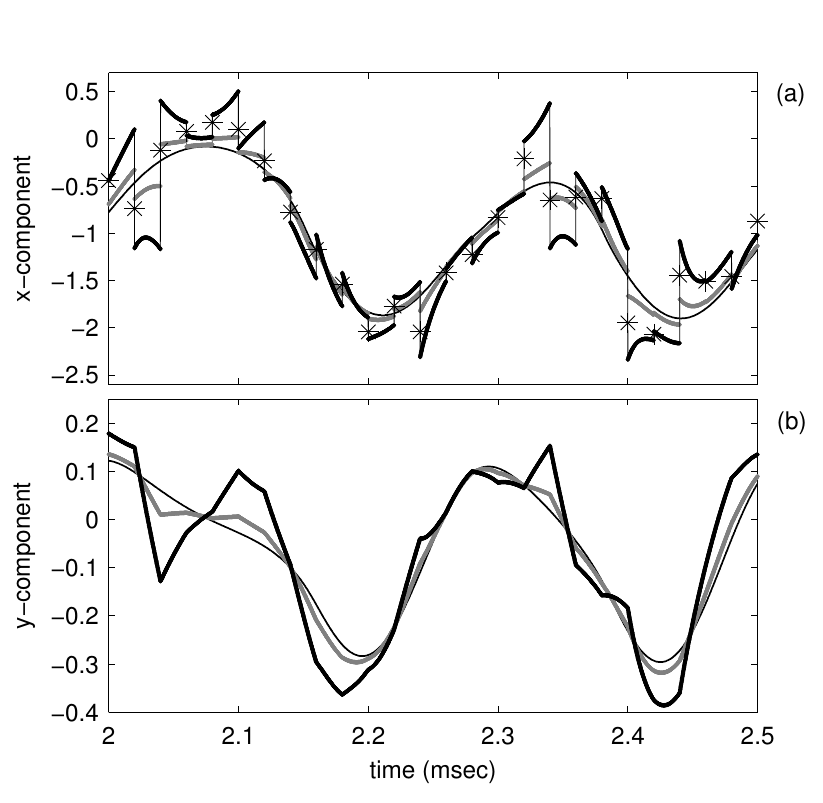}
\caption{Time series of the observed trajectory (thin black), and the
  slave trajectories coupled to observations with $D=10\%$ (thick
  gray) and $40\%$ (thick black) noise with
  $T_{\textrm{obs}}=2.0\times10^{-5} sec$. The noisy observations with
  $D=40\%$ are also shown (stars).}
\label{fig:time_sr_nz}
\end{figure}

To study the effect of noise on synchronization, we added noise to the
$x$-component of observation with $T_{\textrm{obs}}=2.0\times10^{-5}
sec$, which is used in Sec.~\ref{subsec:obs_time}, and repeated the
same experiments. We have not performed any experiments with noisy
continuous time observations, since we are only considering
deterministic models.

In Fig.~\ref{fig:rms_nz} (a) and (b), the solid lines represent the
RMS errors in $x$-component and total error respectively when added
noise intensity is $D=2\%$. In this case, we got synchrony in the
range of $0.66<k_x<2.1$, and the model started desynchronizing around
$k_x=2.1$ as shown in Figs.~\ref{fig:rms_nz}. So the noisy
observations squeeze the k range further. With increase in $D$ overall
range remained almost same. In Figs.~\ref{fig:rms_nz} solid, dashed,
dotted, dashed-dotted, and gray lines are the RMS errors when
$D=10\%$, $20\%$, $30\%$ and $40\%$ respectively.

Effect of the noise is also clear from the time series
plots. Fig.~\ref{fig:time_sr_nz} shows the time series of the
observation (thin solid line) and slave trajectories for $D=10\%$ (thick gray)
 and $40\%$ (thick solid line). They show that the deviation of the
slave trajectories from the observed one increases with increasing
noise level. But $z$-component deviate lesser than $x$ and $y$-component (not shown).
It is also observed that the RMS errors increases almost linearly
(not shown here) with increases in $D$.

\subsection{Prediction with truths and noisy observation}

\begin{figure}[t]
\includegraphics[width=8.5 cm]{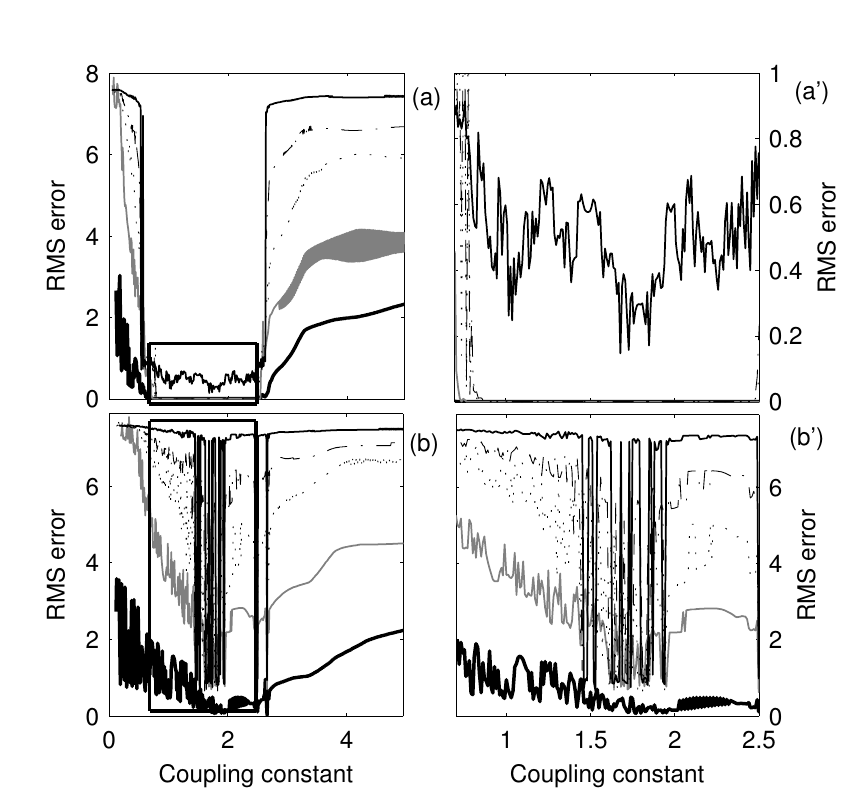}
\caption{Prediction RMS errors in the $x$-component for different
  $t_{fh}$ when $x$ was observed without noise with
  $T_{\textrm{obs}}=2.0\times10^{-5}$ (top) and
  $T_{\textrm{obs}}=5.0\times10^{-5}$ (bottom). Plots in the right
  column are the zoomed version of the boxed regions of the left
  plots.  In each plot, the thick solid, gray, dotted, dash-dot, and
  thin solid lines show the prediction RMS errors when $t_{fh}=
  \frac{1}{10} t_\lambda$, $t_\lambda$, $5t_\lambda$, $10t_\lambda$
  and $50t_\lambda$ respectively.}
\label{fig:rms_x_predict}
\end{figure}

\begin{figure}[t]
\includegraphics[width=8.5 cm]{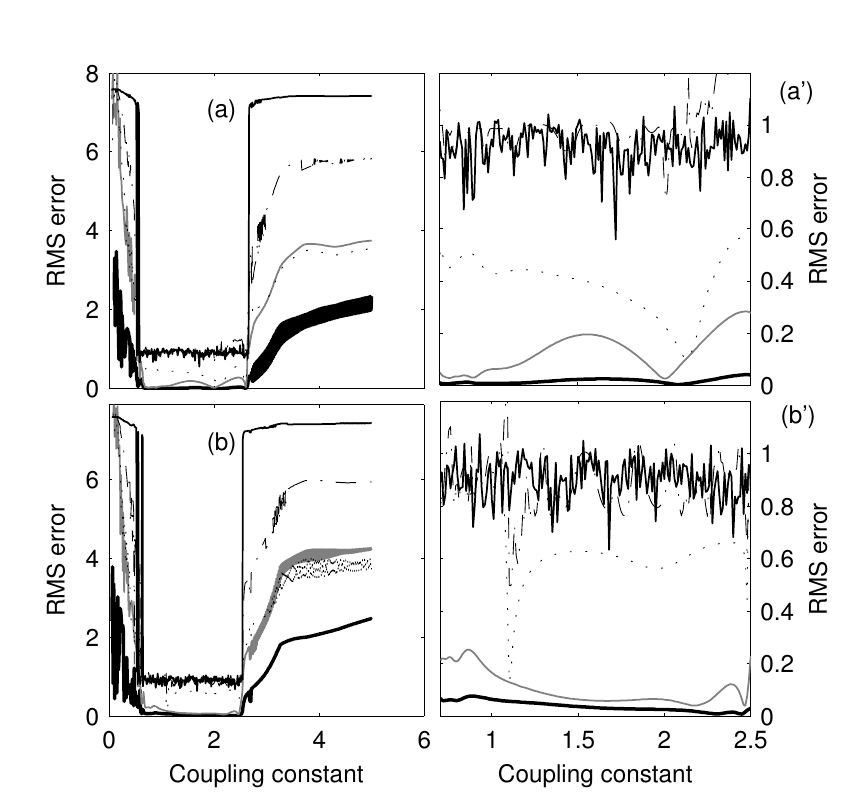}
\caption{Prediction RMS errors in the $x$-component for different
  $t_{fh}$ when $x$ was observed with noise levels $D=6\%$ (top) and
  $D=10\%$ (bottom) with $T_{\textrm{obs}}=2.0\times10^{-5}$. Plots in
  the right column are the zoomed version of the boxed regions of the
  left plots.  In each plot, the thick solid, gray, dotted, dash-dot,
  and thin solid lines show the prediction RMS errors when $t_{fh}=
  \frac{1}{10} t_\lambda$, $t_\lambda$, $5t_\lambda$, $10t_\lambda$
  and $50t_\lambda$ respectively.}
\label{fig:predict_noise}
\end{figure}

\begin{figure}[t]
\includegraphics[width=8.5 cm]{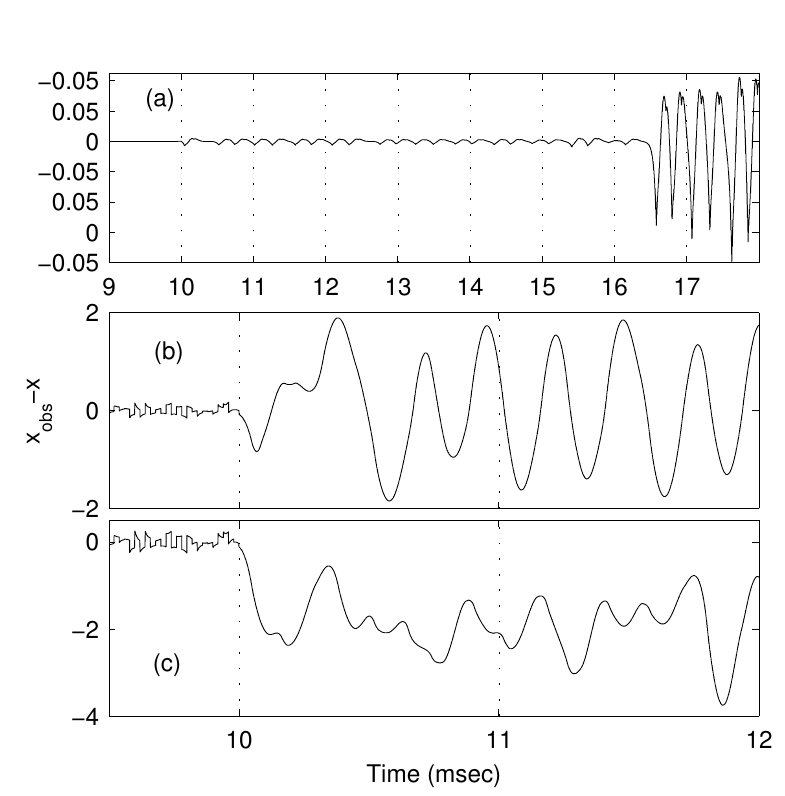}
\caption{Differnce between the observed and the predicted
  trajectories, when the coupling is switched off at $t = 10$
  msec. Top panel (a) is for the case of observations with no noise,
  and the bottom panels (b) and (c) are for the case with noisy
  observations with intensity $D = 6\%$ and $D = 10\%$ respectively.}
\label{fig:predict_traj}
\end{figure}

In this section we consider the use of the slave model for
prediction. This is done in the following manner. We couple the
observations with the slave system through Eqs.~\eqref{eqn:sync} for a
long time period, in particular, until $t_{coup} = 10$ msec in our
numerical experiments. Then we switch off the coupling, i.e., use
Eqs.~\eqref{eqn:master} for the slave beyond this time. The RMS errors
described below are calculated over varying time periods, which are
called the forecast horizons $t_{fh}$. We present the results for
$x$-observations with $T_{\textrm{obs}}=2.0\times10^{-5}$ and
$T_{\textrm{obs}}=5.0\times10^{-5}$ since larger $\tobs$ do not show
any synchronization.

In Fig.~\ref{fig:rms_x_predict} the prediction RMS errors in
$x$-component has been shown for the range of $0<k_x<5$. The $y$- and
$z$- components follow the same qualitative trend. Thick-solid, gray, dotted, dash-dotted and thin solid lines
in Fig.~\ref{fig:rms_x_predict} are the RMS errors for
$t_{fh}=\frac{1}{10} t_\lambda, t_\lambda, 5t_\lambda, 10t_\lambda$,
and $50t_\lambda$ respectively, where boxed regions of
Fig.~\ref{fig:rms_x_predict} (a) has been zoomed in
Fig.~\ref{fig:rms_x_predict} (a$'$).  We see that for
$t_{fh}=\frac{1}{10} t_\lambda, t_\lambda, 5t_\lambda, 10t_\lambda$
slave system predicts well, i.e., the RMS errors are small, but when
$t_{fh}>10t_\lambda$ prediction diverges from true
trajectory. Figs~\ref{fig:rms_x_predict} (a$'$) [thin solid line]
show that RMS error at $t_{fh}= 50t_\lambda$ is significantly
large. Fig.~\ref{fig:predict_traj} (a) show that slave model becomes
completely uncorrelated with the master around $t_{fh}= 40t_\lambda$.

Another important observations from our numerical experiments is that
the forecast horizon for which the slave shows small RMS error with
respect to the master is highly dependent on the state of the system
at which the coupling is turned off. This is because the uncertainty
growth rate is different on different parts of the
attractor.~\cite{pla:Ziehmann}

When noisy observations are coupled with the slave system, the
divergence of the slave from the master is rapid once the coupling is
switched off. Fig.~\ref{fig:predict_noise} shows errors in
$x$-components when noise with intensity $D=6\%$ (top panels) and
$10\%$ (bottom panels) is added to the observations. We have seen that
for $t_{fh} \le \frac{1}{5}t_\lambda$ (not shown in figure), there is
a range of coupling constants for which the prediction RMS error is
small enough to give good prediction. For larger $t_{fh}$, the slave
model become uncorrelated with the master and predictions cannot be
obtained.  This is the case for both cases of the observations with
$6\%$ and $10\%$ noise. Divergence is also clear from the difference
of $x_{obs}-x$ time series of with $6\%$ of noise
[Fig.~\ref{fig:predict_traj} (b)] and $10\%$ of noise
[Fig.~\ref{fig:predict_traj} (c)] respectively.

\section{Conclusion}
\label{sec:conclude}

This paper is devoted to the study of discrete-time unidirectional
synchronization, with specific emphasis on aspects which are of
relevance to data assimilation problems. In particular we consider the
observations of a system, in this case the chaotic Chua system of
Eqs.~\eqref{eqn:master}, to be the master coupled unidirectionally to
the slave system of Eqs.~\eqref{eqn:sync} and examine how the slave
system synchronizes with the master as we change various
characteristics of the observations.  The main discussion is focused
on studying the effects of changing observational period, i.e. the
time between the observations $\tobs$, and observational noise levels
$D$. Some of the main results as well as directions for future
research which we are currently pursuing are discussed below.

We obtained synchronization with continuous time observations of the
$x$- and $y$-components of the system, but not with $z$-component
observations. When discrete observations with finite
$T_{\textrm{obs}}$ were used, Chua model showed synchrony for a finite
range of coupling constants, and this range decreases with increase in
$T_{\textrm{obs}}$ and noisy observations. The study also shows that
the system shows synchrony only when the observations were taken with
$T_{\textrm{obs}}$ much less than $t_{\lambda}$ (almost
$\frac{1}{7}t_{\lambda}$) while for larger $\tobs$ there is no
synchronization between the master and the slave.

In many cases of practical interest in earth sciences, data
assimilation is used as a tool to improve prediction of the observed
as well as unobserved components of the system. With this in mind, we
also study the prediction errors over different forecast horizons
after the coupling between the observations and the slave is turned
off.  We see that in the cases when synchronization errors are small,
the prediction errors even for forecast horizons far beyond the
Lyapunov timescale are small. This happens only for the case of
observations which are not noisy. When using observations even with a
very small noise, the slightly higher synchronization errors coupled
with chaotic nature of the system lead to loss of predictive power
over forecast horizons which are a fraction of the Lyapunov timescale.
These results indicates that the association of the Lyapunov exponent
with predictability horizons has limited use.\cite{pla:Ziehmann} These
results also show the limitations of discrete-time synchronization
with noisy observations and the importance of the study of other data
assimilation techniques which systematically take into account the
noise in the observations.

In synchronization choice of coupling constant $k$ is merely ad hoc
and depends on a particular application. In case of data assimilation
using similar techniques, e.g. nudging\cite{ngp:aurox}, the choice of
$k$ can be interpreted in terms of variational approach. In
particular, Ref.~\onlinecite{ngp:aurox} shows, using the variational
approach, that the optimal coupling constant is inversely proportional
to observational noise covariance: $k \sim R^{-1}$, where $R$ is the
covariance matrix of observational errors. As the range of $k$ on
which model shows synchrony depends mainly on observed component and
noise, we are investigating the possibility that a similar formulation
may also be possible for synchronization to choose the appropriate
$k$. We will also be comparing the synchronization based methods with
data assimilation techniques such as the ensemble Kalman
filter\cite{book:evensen} and Bayesian methods.\cite{physd:apte07}

Another ongoing investigation is to study the performance of these
methods in the chaotic Chua model using real data from the
circuit\cite{pre:singla}. This study will give us an insight about the
choice of coupling constants and observational time period, the
effects of observational noise as well as of the errors in the
numerical model.

\section*{Acknowledgments}

AA and MN would like to thank the Indian National Centre for Ocean
Information Services for financial support through Project
No. INCOIS/93/2007.

\nocite{*}
\bibliographystyle{aipnum4-1}
\bibliography{manus_f}

\end{document}